\documentclass[fleqn,twoside]{article}
\usepackage{espcrc2}
\usepackage[normalem]{ulem}
\usepackage{url}
\urldef{\het}\url{www.het.brown.edu}
\newcommand\email{\begingroup \urlstyle{tt}\Url}
\urldef{\gerry}{\email}{gerry@het.brown.edu}
\usepackage[dvips,usenames]{color}
\usepackage[american]{babel}
\usepackage[latin1]{inputenc}
\usepackage[T1]{fontenc}
\usepackage{xspace}
\usepackage{graphicx}
\usepackage{latexsym}
\usepackage{amsmath,amsfonts,amstext,amssymb,amsbsy,amsopn,amsthm,eucal}
\usepackage{yfonts}[1998/10/03]
\usepackage{dsfont}
\usepackage{wasysym}
\usepackage{wrapfig}
\newcommand{\dslash}{\not{\hbox{\kern-2pt $\partial$}}}
\newcommand{\pslash}{\not{\hbox{\kern-2.3pt $p$}}}
 \newtoks\nslashfraction
 \nslashfraction={.13}
 \newcommand{\nslash}[1]{\setbox0\hbox{$ #1 $}
   \setbox0\hbox to \the\nslashfraction\wd0{\hss \box0}/\box0 }
\newcommand{\plpl}{\raise-2pt\hbox{$\raise3pt\hbox{$_+$}\hskip-6.67pt\raise0.0pt
  \hbox{$^+$}\hskip 0.01pt$}}
\newcommand{\mimi}{\raise-2pt\hbox{$\raise3pt\hbox{$_-$}\hskip-6.67pt\raise0.0pt
  \hbox{$^-$}\hskip 0.01pt$}}
\newcommand{\bo}{\raise-1mm\hbox{\Large$\Box$}}              
\newcommand{\nTH}{{\raise.2ex\hbox{$\displaystyle \bigodot$}\mskip-4.7mu \llap H \;}}
\newcommand{\face}{{\raise.2ex\hbox{$\displaystyle \bigodot$}\mskip-2.2mu \llap
{$\ddot
        \smile$}}}                                      
\newcommand{\mv}[2]{\langle#1|#2\rangle}                  

\newcommand{\leftrightarrowfill}{$\mathsurround=0pt \mathord\leftarrow \mkern-6mu
        \cleaders\hbox{$\mkern-2mu \mathord- \mkern-2mu$}\hfill
        \mkern-6mu \mathord\rightarrow$}
\newcommand{\dvec}[1]{\vbox{\ialign{##\crcr
        \leftrightarrowfill\crcr\noalign{\kern-1pt\nointerlineskip}
        $\hfil\displaystyle{#1}\hfil$\crcr}}}           

\newcommand{\sfrac}[2]{{\vphantom1\smash{\lower.5ex\hbox{\small$#1$}}\over
        \vphantom1\smash{\raise.4ex\hbox{\small$#2$}}}} 
\newcommand{\bfrac}[2]{{\vphantom1\smash{\lower.5ex\hbox{$#1$}}\over
        \vphantom1\smash{\raise.3ex\hbox{$#2$}}}}       
\newcommand{\afrac}[2]{{\vphantom1\smash{\lower.5ex\hbox{$#1$}}\over#2}}    
\newskip\humongous \humongous=0pt plus 1000pt minus 1000pt

\newif\ifdtup

\newfont{\go}{ygoth.tfm scaled 1200}                   
\newfont{\biggo}{ygoth.tfm scaled 3583}                
\newfont{\rope}{cmsy10 scaled 1200}                    
\newfont{\fib}{cmfi10 scaled 1200}
\newfont{\bigfib}{cmfi10 scaled 3583}
\newfont{\funny}{cmff10 scaled 1200}
\newfont{\bigfunny}{cmff10 scaled 3583}
\newfont{\pbk}{pbkd.tfm scaled 1200}
\newfont{\rsfs}{rsfs10.tfm scaled 1200}
\newfont{\bigrsfs}{rsfs10.tfm scaled 2000}
\newfont{\testea}{cmfrak.tfm}
\newfont{\testeb}{dcfrak.tfm}
\newfont{\testec}{schwell.tfm}
\newfont{\tested}{yfrak.tfm}
\newfont{\testef}{yswab.tfm}
\newfont{\bigtestef}{yswab.tfm scaled 3583}
\newfont{\testeg}{yinit.tfm}
\newfont{\testeh}{yinitdd.tfm}
\newfont{\testei}{suet14.tfm}
\newfont{\testej}{pzdr.tfm}
\newfont{\testek}{pzcmi.tfm}
\newfont{\testem}{ccr10.tfm}
\newfont{\testen}{eurm10.tfm}
\newfont{\testeq}{euex10.tfm}
\newfont{\testeo}{wncyr10.tfm}
\newfont{\testep}{msam10.tfm}
\catcode`@=11
\newcommand{\un}[1]{\relax\ifmmode\@@underline#1\else
        $\@@underline{\hbox{#1}}$\relax\fi}
\catcode`@=12

\title{Alternative Numerical Techniques}
\author{G. S. Guralnik\footnote{\gerry} \address[HET]{Brown University, Providence,
    RI. 02912.}, J. Doll \addressmark[HET], R. Easther\address[COL]{Columbia
    University, New York, NY. 10027.},
  P. Emirdag\addressmark[HET], D. D. Ferrante\addressmark[HET], S.
Hahn\addressmark[HET],
  D. Petrov\addressmark[HET] and D. Sabo\addressmark[HET].}
\begin{document}
\begin{abstract}
Two new approaches to numerical QFT are presented.
\end{abstract}
\maketitle
\section{Introduction}
``Traditional'' Monte Carlo (MC) approaches have produced many good
results but they do have well known limitations.  These include
requiring large amounts of computation time as a result of the
necessity to treat fermions very differently from bosons, as well as
problems with fermion multiplicity associated with the lattice.
Moreover, non-positive definite actions, actions with rapid
oscillations including phase transitions and symmetry breaking are
generally not tractable. Finally, highly accurate answers can require
immense resources.

For these reasons we have been working on two alternative approaches.
The first is a tuned Monte Carlo method, which we call Mollified Monte
Carlo (MMC), while the second, involving nested approximations to the
Schwinger--Dyson equations, is called the Source Galerkin method (SG).
These two methods are closely related because MMC is initiated by
starting from a stationary phase point of an action, while the
simplest applications of SG tend to favor such points in initial
iterations.  The talks by Ferrante and Petrov \cite{mmc,tsgm} give a
few more details.

\section{Mollified Monte Carlo}

Unlike normal MC, the integrands for problems approached using MMC are
very carefully controlled before any actual heavy duty calculation
takes place. Regions of high oscillation are identified and smoothed
(mollified) by using simple exponential functions that do not change
the values of any of the integrals but do suppress unimportant
oscillatory contributions. The resulting integrals are evaluated
around stationary points using appropriate importance function
weighting and MC integration. In the cases we have examined this
approach allows accurate evaluation of normally intractable
integrands, including complex integrands. In particular, theories can
be directly evaluated in symmetry breaking or other phases with
comparative ease. MMC still suffers from the traditional problems
associated with fermions (although with the potential for more rapid
convergence of integrals) but does considerably extend the class of
actions and regions of evaluation open to the usual approaches.

\section{Source Galerkin Methods}
Our Source Galerkin method, while requiring less familiar numerical
techniques, is powerful because it is defined on the continuum and
treats fermions, except for anti-commutativity, in the same way as
bosons. The basic idea is simple: We assume that the QFT action is
written with sources $J_i(x)$ for every field $\phi_i(x)$ so that the
amplitude $Z = \mv{+0}{0-}_{J_i}$ satisfies the differential equation,
\begin{equation*}
  \underbrace{ F\Bigl( \frac{\delta}{\delta J_i(x)} \Bigr)
}_{\substack{\text{essentially the}
    \\ \text{field equations}}}\, \mathcal{Z} = 0 \; .
\end{equation*}

A familiar but far from straightforward example is $\phi^4$ scalar QFT,
\begin{equation*}
  (\partial^2 + m^2)\, \phi(x) + g\, \phi^3(x) = J(x) \; ,
\end{equation*}
which becomes (in Euclidean space),
{\small
\begin{equation*}
  \Biggl[(\partial^2_x + m^2)\,\frac{\delta}{\delta J(x)}  +
  g\,\Bigl(\frac{\delta}{\delta J(x)}\Bigr)^3 - J(x)\Biggr] \mathcal{Z}[J] = 0 \; .
\end{equation*}
}
This is a very non-trivial infinite set of coupled differential equations. To get
some
idea of its complexity, examine the one point (0-dim) case,
{\small
\begin{equation*}
  \Biggl[ m^2\, \frac{d}{d J} + g\, \Bigl( \frac{d}{d J} \Bigr)^3 - J \Biggr]
  \mathcal{Z}[J] = 0 \; .
\end{equation*}
} Even this simple equation has three independent solutions and
requires the input of parameters for a full solution \cite{tvbcsde}.
These three solutions can be characterized by their degree of
singularity:
\begin{itemize}
\item {Regular at $g\rightarrow 0$:} consistent with perturbation theory;
\item {$\sim g^{-1/2}$ at $g\rightarrow 0$:} ``symmetry breaking'';
\item {$\sim \exp(\mu^2/4\,g)$ at $g\rightarrow 0$:} ``instanton''.
\end{itemize}
In the path integral language, these solutions correspond to the three
saddle points of the path integral in the complex $\phi$-plane. 

The Schwinger condition,
\begin{equation*}
  \frac{d\, \mathcal{Z}}{d\, g} = \varint\, (dx)\, \biggl(\frac{\delta}{\delta
    J(x)}\biggr)^4\, \mathcal{Z} \; ,
\end{equation*}
is used to stabilize the phase choice \cite{tvbcsde} using SG in iterative
approximations as is discussed in what follows.

A direct approach to solving these is to assume a truncated functional
expansion in the sources and introduce some way to minimize the error.
It is the presence of a method for error control that differentiates
our approach from any older Schwinger-Dyson approximations.
Formally, the correct answer is, {\footnotesize
  \begin{equation*}
    \mathcal{Z} = \exp\biggl\{ A_0 + \varint A_i(x)\, J_i(x) + \varint
    A_{ij}(x,y)\, J_i(x)\, J_j(y) + \dotsm \biggr\}
  \end{equation*}
} The problem is to introduce a procedure to iteratively calculate the
$A_i$'s. In order to do this, we put information we already have about
the field theory (spectral information, relativity, Lehman-Kallen
representations) and make decompositions by assuming
{\small
\begin{flalign*}
  &\mathcal{Z} = \exp\biggl\{a + \varint n\, J(x) + \varint J(x)\, G(x-y)\, J(y)
    \biggr\} + \\
  &\; + \{\text{all possible higher order Green's function terms} \\
  &\text{all constructed from terms involving only functions} \\
  &\text{$G(x-y)$ but with possibly different parameterization.}\}
\end{flalign*}
} The above representation is highly symbolic. To make calculations
tractable we have spanned the solution space involving products of
sources by multiplying these, in turn, with all possible products of
free field propagators with arbitrary weights and masses.  For
example, the quartic source contribution, $ J J G_4JJ$ is represented
by all possible graphs involving only the ``free'' $G(x-y)$ with four
external lines. The $G(x-y)$ can have arbitrary numerical weights and
mass. The result is a very general, very complicated expression
consistent with Lorentz and translation invariance.

The propagator $G(x-y)$ is,
\begin{equation*}
  G(x-y) = \varint dK^2\, d^4k\, \frac{e^{i\, k(x-y)}}{-k^2 + K^2}\, a(K^2)
\end{equation*}
This satisfies space-time restrictions, and the $a(K^2)$ are weight
functions.  When inserted into the field equations the above expansion
yields constraints on $a(K^2)$.  In general, this is too complicated
to directly solve, since this would be tantamount to an exact solution
of a field theory. Finally, there are cross-terms in any equation and
the equations are non-linear! 

We simplify as follows:
\begin{itemize}
\item Truncate the expansion in $J$;
\item Limit the number of masses in each propagator;
\item Limit the number of graphs considered for each $J$.
\end{itemize}
  
We do this in an organized, systematic, way so that after a first
guess more terms can be included so as to iterate answers.

Roughly this works as follows: Approximating yields
$Z^*[J]$ and, of course,
\begin{equation*}
  F\biggl(\frac{\delta}{\delta J}\biggr)\, \mathcal{Z}_{\text{approx}}[J] \neq 0 \; .
\end{equation*}
The idea of Source Galerkin is to require that,
\begin{equation*}
  \varint\, dJ_1\dotsb dJ_n\, F_i(J_1,\dotsc,J_n)\, F\biggl(\frac{\delta}{\delta
    J}\biggr)\, \mathcal{Z}_{\text{approx}}[J] = 0 \; ,
\end{equation*}
so that $\mathcal{Z}_{\text{approx}}[J]$ satisfies the field equations
``on the average''. An appropriate number of $F_i$ are picked so that
all the undetermined weights in $Z^*[J]$ are determined.

In general, these equations are non-linear, and must be solved in a
very careful and systematic manner. Theorems for simpler Galerkin
approaches promise convergence and we conjecture that the same is true
for QFT.  We have applied this approach to many models (with amazing
accuracy when a check is available) but the stability and convergence
must be confirmed.  Higher iterations are simple in principle, but
introduce computational difficulties.  To understand these we have
re-examined perturbation theory using variants of our numerical
techniques [4] and also re-analyzed trivial 0-dimensional models for
higher order Galerkin expansion [2].

\section{Perturbation Theory}
The usual graphical rules apply (not our extended
``exact'' rules). Application of our ideas have lead to a new way to
numerically calculate graphs \cite{fefd}. The major ingredients of our
new procedure are the Sinc-function expansion:
\begin{flalign*}
  S_k (h,x) &\equiv \frac{\sin(\pi\, (x - k\, h)/h)}{\pi\, (x - k\, h)/h}\; ,\quad
    k\in\mathbb{Z} \; ;\\
  \intertext{and the cutoff propagator}
  G_{\Lambda}(x) &= \varint\frac{d^4p}{(2\, \pi)^4}\; \frac{e^{i\, p\, x}}{p^2 +
    m^2}\, e^{-p^2/\Lambda^2} \; .
\end{flalign*}
Using the Sinc expansion gives [4]
\begin{flalign*}
  G_{\Lambda,h}(x) &= \frac{m^2\, h}{(4\, \pi)^2}\, \sum_{k=-\infty}^{\infty}p(k)\,
    \exp\biggl[-\frac{m^2\, x^2}{4\, C(k)}\biggr] \\
  C(k) &= e^{k\, h} + \frac{m^2}{\Lambda^2} \;,\quad p(k) = \frac{e^{k\,h - e^{k\,
    h}}}{C^2(k)} \; .
\end{flalign*}
    
Typically, this can be approximated to very high accuracy (1 part in
$10^{16}$) with fewer than 100 terms in the sum and similar statements
hold for fermion propagators. Using these propagators, graphs can be
reduced to multi-dimensional sums which can then be quickly (relative
to the analogous Monte Carlo integral) calculated to high precision.
We believe this may be a powerful way to check high order magnetic
moment calculations, which are currently of great interest.  Technical
issues with accuracy using an ``auto'' renormalization based on
approaches used in lattice gauge theory scheme have slowed us down,
but we hope to release some more computations emphasizing the power of
this method in the near future.  Moreover, the ease and speed of these
perturbative calculations gives us hope that we can iterate Source
Galerkin calculations to moderate order with relatively small amounts
of computer time.
\section{Acknowledgments}
DDF, DP and GSG wish to acknowledge support by \textsf{DOE} grant
\textsf{DE-FG02-91ER40688 - Task D}\/, JDD and DS wish to acknowledge support from
the
\textsf{National Science Foundation} through awards \textsf{CHE-0095053} and
\textsf{CHE-0131114}\/.
\vspace{1pc}
\end{document}